\documentclass[aps,pra,twocolumn,showpacs,groupedaddress]{revtex4-1}
\usepackage{amsmath}
\DeclareMathOperator{\sech}{sech}
\usepackage{amsthm}
\usepackage[english]{babel}
\usepackage[T1]{fontenc}
\usepackage[latin1]{inputenc}
\usepackage{graphicx}
\usepackage{amssymb}
\usepackage{color}

\def\r {{\bf r}}

\begin{document}
\title{Single shot simulations of dynamic quantum many-body systems}


\author{Kaspar Sakmann$^{1}$ and Mark Kasevich$^1$}

\affiliation{$^1$ Department of Physics, Stanford University, Stanford, California 94305, USA}

\begin{abstract}
{\bf The single-particle density is the most basic quantity 
that can be calculated from a given many-body wave function.
It provides the probability to find a particle at a given position 
when the average over many realizations of an experiment is taken.
However, the outcome of single experimental shots of ultracold atom experiments 
is determined by the $N$-particle probability density. This difference can lead to surprising results.
For example, independent Bose-Einstein condensates (BECs) 
with definite particle numbers form interference fringes even though no 
fringes would be expected based on the single-particle density \cite{JavYoo,Ket,CasDal,Joerg06}.
By drawing random deviates from the $N$-particle probability density 
single experimental shots can be simulated from first principles \cite{JavYoo,CasDal,Lew09}.
However, obtaining expressions for the $N$-particle probability density 
of realistic time-dependent many-body systems has so far been elusive. 
Here, we show how single experimental shots of general 
ultracold bosonic systems can be simulated based on numerical solutions of 
the many-body Schr\"odinger equation. We show how full counting distributions 
of observables involving any number of particles can be obtained 
and how correlation functions of any order can be evaluated. As examples
we show the appearance of interference fringes in interacting independent BECs,
fluctuations in the collisions of strongly attractive BECs, 
the appearance of randomly fluctuating vortices in rotating systems 
and the center of mass fluctuations of attractive BECs in a harmonic trap. 
The method described is broadly applicable to bosonic many-body systems 
whose phenomenology is driven by information beyond what is 
typically available in low-order correlation functions.}
\end{abstract}

\maketitle
Let us briefly outline how single experimental shots can be simulated
from a general many-body wave function $\Psi$. 
The probability to find $N$ particles
at positions $\r_1,\dots,\r_N$ in a many-body system 
is determined by the $N$-particle probability distribution 
$P(\r_1,\dots,\r_N)=\vert\Psi(\r_1,\dots,\r_N)\vert^2$. 
In experiments on ultracold bosons snapshots of the positions of the particles are taken 
and single experimental shots sample the $N$-particle probability distribution.
This distribution is high-dimensional and sampling it directly from a given $N$-boson wave function 
is hopeless. However, it can be rewritten as a product of conditional probabilities
\begin{equation}\label{Pr}
P(\r_1,\dots,\r_N)=P(\r_1)P(\r_2\vert\r_1)\times\dots\times P(\r_N\vert\r_{N-1},\dots,\r_1),
\end{equation}
where e.g. $P(\r_2\vert\r_1)$ denotes the conditional probability to find a particle at 
$\r_2$ given that another particle is at $\r_1$.   
By drawing $\r_1$ from $P(\r)$, $\r_2$ from $P(\r\vert\r_1)$, $\r_3$ from $P(\r\vert\r_2,\r_1)$, etc., 
one random deviate of $P(\r_1,\dots,\r_N)$ is generated.  
Obtaining the conditional probabilities in (\ref{Pr}) 
is a formidable combinatorial problem though, even for special cases \cite{JavYoo,Lew09}. 
Here, we provide a general algorithm to simulate single shots from 
any given $N$-boson wave function $\vert \Psi\rangle=\sum_{\vec n} C_{\vec n} \vert\vec n\rangle$, where 
$\vert \vec n\rangle=\vert n_1,\dots,n_M\rangle$ are configurations constructed by distributing $N$ bosons
over $M$ orbitals $\phi_i$.
We apply this algorithm to many-body states obtained by solving the time-dependent many-body Schr\"odinger equation
numerically using the multiconfigurational time-dependent Hartree for bosons method (MCTDHB) \cite{MCTDHB,Mapping,Package}. 
This combination of many-body Schr\"odinger dynamics and sampling of the $N$-particle probability 
allows us to simulate single experimental shots from first principles in realistic settings, 
see Methods for the algorithm and details.
\begin{figure}[ht]
\includegraphics[width=6cm, angle=-90]{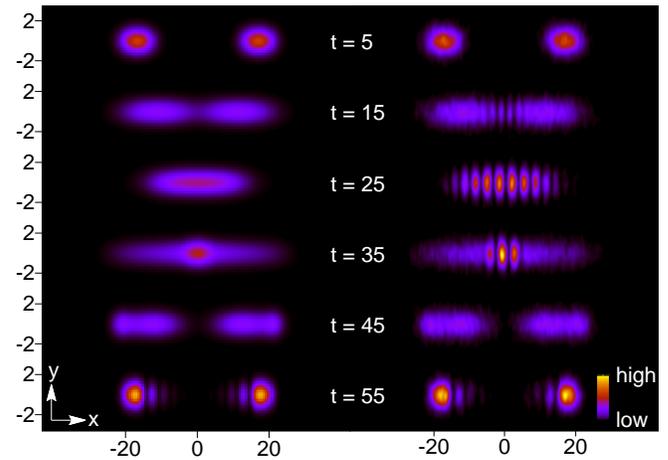}
\caption{Interference of independent interacting condensates.
Two independent, repulsively interacting condensates collide in an elongated trap. 
Shown is the single-particle density (left column) and random deviates of the 
$N$-particle density (right column) at different times.
In the overlap region interference fringes show up in the $N$-particle density, but not in the 
single-particle density. The results are obtained by solving the many-body Schr\"odinger equation in two spatial dimensions. 
Parameter values: $N=10000$ bosons. Interaction strength $\lambda=4.95$. See text for details. All quantities shown are dimensionless.}\label{F1}
\end{figure}

It is instructive to briefly review Bose-Einstein condensation. 
A many-boson state is condensed if its reduced single-particle density 
matrix 
has exactly one nonzero eigenvalue $\rho_i$ of order $N$ \cite{Pen56}. The eigenvalues $\rho_i$ are known as natural occupations, 
the eigenvectors
as natural orbitals. The BEC is fragmented if more than one eigenvalue $\rho_i$ is of order $N$ \cite{Noz,MCHB}, see Methods for details.
Fully condensed states, i.e. states with $\rho_1=N$,  are of the form $\phi(\r_1)\phi(\r_2)\times\dots\times\phi(\r_N)$. 
(\ref{Pr})  then becomes a trivial product of independent, identical probability distributions, 
and there are no correlations between particles. For instance, Gross-Pitaevskii (GP) mean-field states are of this form. 
Any other state, in particular fragmented states, exhibit correlations and vice versa any correlated state is to some degree fragmented.
We will now show how fragmented BECs lead to macroscopically fluctuating outcomes in single shots.

In the following we use dimensionless units $\hbar=m=1$ and solve the time-dependent 
many-body Schr\"odinger equation $i\frac{\partial}{\partial t}\vert \Psi \rangle=\hat H\vert\Psi\rangle$ using the MCTDHB method \cite{MCTDHB,Mapping,Package}. 
Here, 
\begin{equation}\label{Ham}
H=\sum_{i=1}^N -\frac{1}{2}\frac{\partial^2}{\partial \r_i^2}+ V(\r_i)+\lambda_0\sum_{i<j}\delta_\epsilon(\r_i-\r_j)
\end{equation}
denotes a general many-body Hamiltonian in $D$ dimensions with an external potential $V(\r)$ and a 
regularized contact interaction  $\delta_\epsilon(\r)=(2\pi\epsilon^2)^{-D/2}e^{-\r^2/2\epsilon^2}$. 
We parameterize the interaction strength by the mean-field parameter $\lambda=\lambda_0(N-1)$, see Methods for details.

Let us begin with an example of two interfering, independent condensates of 
$N=10000$ bosons in an elongated trap with tight harmonic confinement along the $z$ direction such that we can work in $D=2$ dimensions and $\r=(x,y)$.
We use $V(\r)=V_x(x)+V_y(y)+V_g(x)$ as an external potential, where $V_{x}(x)$ and $V_{y}(y)$ are harmonic traps 
and $V_g(x)$
is an additional Gaussian potential that flattens the bottom of the trap along the $x$-direction. 
As an initial state  we use two independent condensates, 
each of which is the  mean-field ground state (corresponding to $M=1$ in the MCTDHB formalism)
of $N/2$ bosons of the displaced traps $V_\pm(\r)=V(x \pm d, y)$ with $d=18.6$ in harmonic oscillator units of the $y$-direction 
at an interaction strength $\lambda=4.95$.
The initial state $\vert \Psi(0)\rangle=\vert N/2,N/2\rangle$ is fragmented with $\rho_1=\rho_2=N/2$.   
We then solve the time-dependent many-body Schr\"odinger equation for $\vert\Psi(0)\rangle$ using $M=2$ orbitals.
Fig. 1 shows the single-particle density as well as random deviates of the $N$-particle density at different times. 
The two condensates accelerate towards each other, 
collide and separate again. During the collision interference fringes 
appear in deviates of the $N$-particle density at locations that fluctuate randomly from shot to shot, 
but not in the single-particle density. This is also expected based on simplified models \cite{JavYoo,CasDal}. However,  
here this result follows directly from the solution of the many-body Schr\"odinger equation.
The interparticle interaction is weak here; interaction effects only 
become visible as ripples in the density after the collision and the natural occupations remain practically constant  all along.

\begin{figure}
\includegraphics[width=6cm, angle=-90]{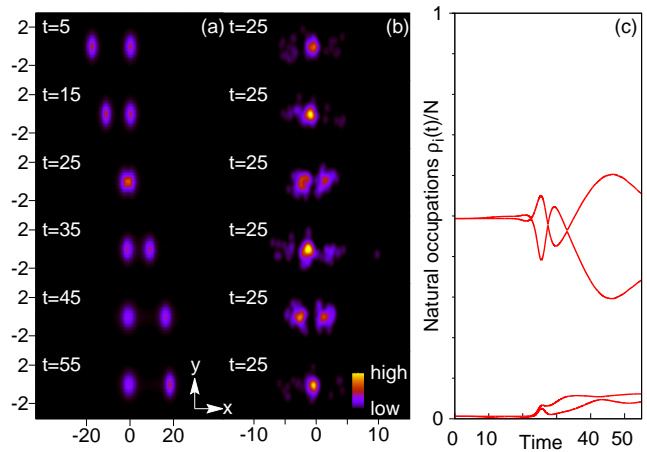}
\caption{Collision of independent attractively interacting condensates. Two independent attractively
interacting condensates collide in an elongated trap in two spatial dimensions. 
(a) Single-particle density at different times. The condensates approach each other without spreading and bounce off one another. 
(b) Random deviates of the $N$-particle density at the time of the collision. 
Correlations lead to either a single strongly localized density maximum containing practically all particles or two 
smaller maxima containing about half the particles each. 
(c) Fragmentation of the condensate as a function of time. The initial state is two fold fragmented with 
$\rho_1/N=\rho_N/N=49.4\%$. During the collision two additional natural occupations become significantly occupied
and the system can no longer be separated into two independent condensates. 
Parameter values: $N=100$ bosons. Interaction strength $\lambda=-5.94$. See text for details. All quantities shown are dimensionless.}\label{F2}
\end{figure}

We now go one step further and investigate collisions between strongly attractive independent condensates in the same trap. 
For this purpose we use $N=100$ bosons at an interaction strength $\lambda=-5.94$ which is about $2\%$ above the 
threshold for collapse of the GP mean-field ground state in this trap. For the initial state we first compute the many-body ground 
state of fifty bosons using two orbitals and imaginary time-propagation. This ground state is highly condensed, $\rho_1/N=98.7\%$. The initial state is then 
taken as the symmetrized product of the ground state and a displaced copy of it located at $\r=(-d,0)$. 
Thus, the initial state has natural occupations $\rho_1/N=\rho_2/N=49.4\%$ and $\rho_3/N=\rho_4/N=0.6\%$.
We then propagate this initial state using $M=4$ orbitals.

Fig. 2 (a) shows the single particle-density at different times. 
The condensates approach each other without spreading 
significantly, collide and separate again. During the collision the single-particle density exhibits two maxima, 
the condensates seem to bounce off each other. However, single shots at the time of the collision reveal a different result, see Fig. 2 (b). 
In about half of all shots a strongly localized density maximum is visible, whereas in the other half two smaller well separated
maxima appear. We stress that at no point any type of (possibly random) phase relationship between the colliding parts 
was assumed. 
In fact, for independent condensates the assumption of a preexisting, but random relative phase is at variance with quantum mechanics \cite{MulLal}.
The macroscopic fluctuations in the outcomes follow  directly from the intrinsic correlations of the many-body state.
Fig. 2 (c) shows the natural occupations of the system. As long as the 
condensates are far apart, the natural occupations remain close to their initial values. 
However, during the collision two additional natural orbitals become occupied indicating a buildup of 
even stronger correlations. As a consequence after the collision the system can no longer be separated  
into two independent condensates.

\begin{figure}
\includegraphics[width=6cm, angle=-90]{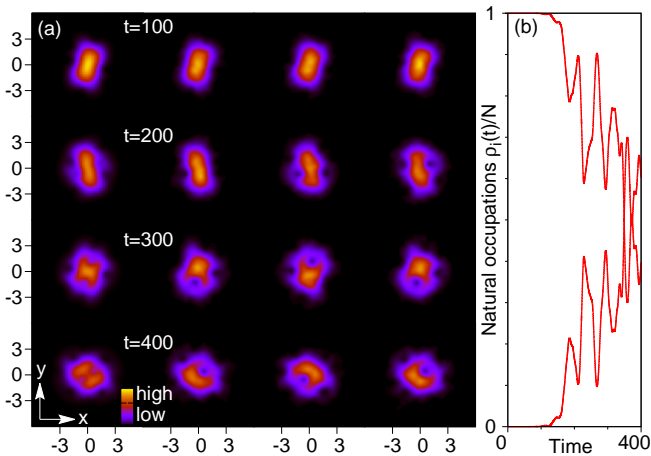}
\caption{Fluctuating vortices. A repulsive condensate in the ground state of a harmonic trap 
is stirred  by a rotating potential in two spatial dimensions. 
Over the course of time the system fragments and vortices appear at random positions
in single shots. (a) First column: single-particle density at different times.
Second to fourth column: single shots at the same times.
(b) Fragmentation of the condensate as a function of time. 
Starting from a condensed state, the system of bosons fragments as it is stirred.
While the system is condensed single shots and the single-particle density look alike. 
When the system is fragmented vortices appear at random positions. 
Parameter values: $N=10000$. Interaction strength: $\lambda=17$. See text for details.
All quantities shown are dimensionless.}\label{F3}
\end{figure}

In the previous two examples already the initial states were fragmented. 
We now turn to a system where fragmentation builds up dynamically.
Stirring a BEC can lead to fragmentation and vortex nucleation 
that cannot be explained within the mean-field framework of quantized vortices \cite{Lew09, Axel}.
Consider the ground state of a repulsively interacting BEC of $N=10000$ bosons in a pancake shaped trap with $\omega_x=\omega_y=1$
at an interaction strength $\lambda=17$. We compute the many-body ground state using $M=2$ orbitals
which is practically fully condensed with $\rho_1/N=99.98\%$. We then switch on a time-dependent 
stirring potential $V_s(\r,t)=\frac{1}{2}\eta(t)[x(t)^2-y(t)^2]$ that imparts angular momentum onto the BEC.
Here $x(t)$ and $y(t)$ vary harmonically 
and the amplitude $\eta(t)$ is linearly ramped up from zero to a finite value and back down, see Methods for details. 
Fig. 3 (a) shows the density  together with single shots at different times. The evolution of the natural occupations is shown in Fig. 3 (b). 
While the system is condensed, single shots reproduce the single-particle density. 
Over the course of time an additional natural orbital becomes occupied and 
the BEC becomes correlated. As correlations build up the outcome of single shots 
fluctuates more and more and vortices appear at random locations in every single shot. 
This is in stark contrast to mean-field theory, where due to the lack of correlations 
vortices always appear at the same location.

\begin{figure}
\includegraphics[width=6cm, angle=-90]{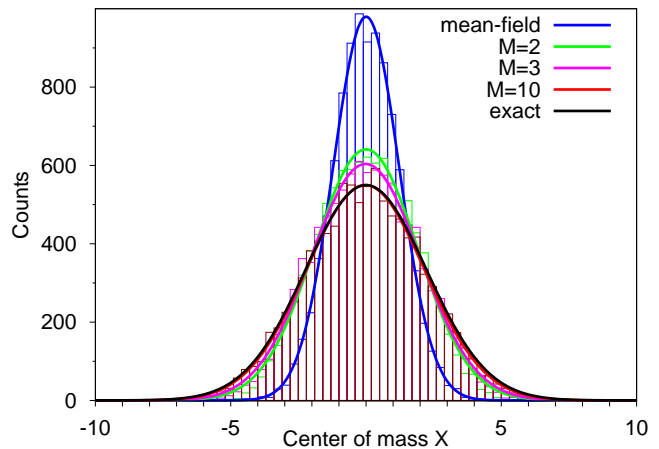}
\caption{Full counting distribution of the center of mass operator. 
Shown are $10000$ random deviates of the center of mass operator of the ground 
state of an attractively interacting condensate in one spatial dimension. 
The center of mass fluctuations of the mean-field result (blue)
are significantly smaller than those of the many-body results where the bosons are allowed to occupy 
$M=2,3,10$ (green, magenta, red) orbitals. The $M=10$ result coincides with the exact analytical one (black). 
Parameter values: $N=10$ bosons; interaction strength $\lambda=-0.423$, trap frequency $\omega_x=1/100$. 
All quantities shown are dimensionless.}\label{F4}
\end{figure}

As a last example let us show how full distribution functions of $N$-body operators
can be evaluated by simulating single shots. Consider the ground state of $N$  attractively interacting
bosons in a harmonic trap, $\omega_x=1/100$, in one dimension, i.e. $D=1$ and $\r=x$. 
The exact wave function of the center of mass coordinate $X=\frac{1}{N}\sum_i{x_i}$ of the many-body ground state 
is given by a Gaussian $\Psi_{mb}(X)=(\sqrt{\pi}X_{mb})^{-1/2}e^{-X^2/2 X_{mb}^2}$ with $X_{mb}=1/\sqrt{N\omega_x}$ \cite{KohnThm}.
On the other hand, the mean-field ground state is uncorrelated and hence its center of mass width
is given by  $X_{mf}=\sigma_{mf}/\sqrt{N}$, where $\sigma^2_{mf}=\langle\phi_{mf}\vert x^2 \vert \phi_{mf}\rangle$ 
is the variance of the mean-field orbital $\phi_{mf}$, see Methods.
In the limit of a weak trap,  $\omega_x\rightarrow 0$,  the mean-field solution approaches a soliton with $\sigma_{mf}=\pi/(\sqrt{3}\vert\lambda\vert)$.
Thus, for sufficiently strong attractive interaction $X_{mb}$ exceeds $X_{mf}$.
We compute the ground state of $N=10$ bosons at an 
interaction strength $\lambda=-0.423$ using imaginary time-propagation for different numbers  of orbitals.  
From the obtained ground states we generate $10000$ random deviates of the center of mass coordinate.
Fig. 4 shows fits to the obtained histograms of the center of mass deviates together with 
the exact center of mass distribution. The many-body result for $M=10$ orbitals is indistinguishable from the exact one
and significantly broader than the mean-field ($M=1$) result. 
In the present example the many-body correlations are the cause for the onset of the delocalization of the ground state.

\section*{Methods}
\subsection*{Bose-Einstein condensation.}
For an $N$-boson state $\vert \Psi\rangle=\sum_{\vec n} C_{\vec n}(t) \vert\vec n\rangle$ 
and a bosonic field operator $\hat\Psi(\r)=\sum_j\hat b_j \phi_j(\r)$
the reduced single-particle density matrix is defined as
\begin{equation}
\rho^{(1)}(\r\vert \r')=\langle \Psi \vert  \hat\Psi^\dagger(\r')\hat\Psi(\r) \vert \Psi\rangle =\sum_{i,j} \rho_{ij} \phi_i^\ast(\r')\phi_j(\r)
\end{equation}
with $\rho_{ij}=\langle\Psi\vert \hat b_i^\dagger \hat b_j\vert \Psi\rangle$. By diagonalizing $\rho_{ij}$ 
one obtains $\rho^{(1)}(\r\vert \r')=\sum_i \rho_i \phi^{NO}_i(\r) \phi^{NO \ast}_i(\r')$.
The eigenvalues $\rho_1\ge\rho_2\ge\dots$ are known as natural occupations, the eigenvectors $\phi^{NO}_i(\r)$ as natural orbitals.
If there is only one eigenvalue $\rho_1={\mathcal O}(N)$ the BEC is condensed \cite{Pen56}, if more than one $\rho_i={\mathcal O}(N)$ 
the BEC is fragmented \cite{Noz,MCHB}.
The diagonal $\rho(\r)\equiv\rho^{(1)}(\r\vert \r'=\r)$ is the single-particle density of the $N$-boson wave function.

\subsection*{Single Shot Algorithm.}
Here we show how  single shots can be simulated from a general $N$-boson wave function 
expanded in $M$ orbitals $\vert \Psi\rangle=\sum_{\vec n} C_{\vec n} \vert\vec n\rangle$, where 
$\vert \vec n\rangle=\vert n_1,\dots,n_M\rangle$ and $\sum_{i=1}^M n_i=N$. Special cases (for $M=2$) have been treated  in earlier works \cite{JavYoo,Lew09}.
The goal is to draw the positions $\r_1,\dots,\r_N$ of $N$ bosons from the probability distribution
$P(\r_1,\dots,\r_N)$. We achieve this by evaluating the conditional probabilities in (\ref{Pr}). 
For this purpose we define reduced wave functions 
\begin{equation}\label{Psik}
\vert \Psi^{(k)} \rangle =
\begin{cases}
\vert\Psi\rangle, & \text{if }k=0 \\
{\mathcal N}_k\hat\Psi(\r_k)\vert\Psi^{(k-1)}\rangle, & \text{if }k=1,\dots,N-1
\end{cases}
\end{equation}
of $n=N-k$ bosons with normalization constants ${\mathcal N}_k$. The respective single-particle densities are given by
$\rho_k(\r)=\langle\Psi^{(k)}\vert \hat\Psi^\dagger(\r)\hat\Psi(\r)\vert \Psi^{(k)}\rangle$ and ${\mathcal N}_k=\rho_{k-1}(\r_k)^{-1/2}$.
The first position $\r_1$ is  drawn from $P(\r)=\rho_0(\r)/N$. 
Assuming that 
positions $\r_k, \dots,\r_{1}$ have already been drawn,
the conditional probability density for the next particle $P(\r\vert \r_{k},\dots,\r_1)= P(\r,\r_{k},\dots,\r_1)/P(\r_{k},\dots,\r_1)$  is given by 
\begin{equation}\label{Pproprho}
P(\r\vert \r_{k},\dots,\r_1)\propto\rho_k(\r),  
\end{equation}
since $P(\r_k,\dots,\r_1)$ is a constant. 
The problem is thus reduced to obtaining the wave function $\vert\Psi^{(k)}\rangle=\sum_{\vec n} C_{\vec n}^{(k)}\vert \vec n\rangle$ 
from the wave function $\vert\Psi^{(k-1)}\rangle=\sum_{\vec n} C_{\vec n}^{(k-1)}\vert \vec n\rangle$,
where the sums over run over all configurations of $n$ and $n+1$ bosons, respectively.
Defining $\vec n^q=(n_1,\dots,n_q+1,\dots,n_M)$ one finds from (\ref{Psik})
\begin{equation}\label{cnk}
C_{\vec n}^{(k)}={\mathcal N}_k \sum_{q=1}^M \phi_q(\r) C_{\vec n^q}^{(k-1)} \sqrt{n_q+1}
\end{equation}
Using (\ref{cnk}) in a general $M$ orbital algorithm requires  
an ordering of the  $n+M-1 \choose n$ configurations $\vert \vec n \rangle$  for all particle numbers $n=1,\dots,N$. 
Combinadics \cite{Mapping} provide such an ordering by associating the index
\begin{equation}\label{Jind}
J(n_1,\dots,n_M)=1+\sum_{i=1}^{M-1} {n+M-1-i-\sum_{j=1}^i n_j \choose M-i}
\end{equation}
with each configuration $\vert\vec  n\rangle$.
Using (\ref{Jind}) all
coefficients $C_{\vec n}^{(k)}$  can then be obtained  by evaluating the sums in (\ref{cnk}) and 
${\mathcal N}_k$  is determined by normalization. Using the coefficients $C_{\vec n}^{(k)}$ we evaluate $\rho_k(\r)$ and by means of  
(\ref{Pproprho}) we then draw $\r_{k+1}$  from $P(\r\vert\r_k,\dots,\r_1)$. 
This concludes the algorithm to simulate single shots.
It is now easy to see that also correlation functions of arbitrary order can be evaluated. 
By realizing that 
\begin{equation}
\langle \Psi\vert \hat\Psi^\dagger(\r_1)\dots \hat \Psi^\dagger(\r_k)\hat\Psi(\r_k)\dots\hat\Psi(\r_1)\vert\Psi\rangle=\prod_{j=1}^{k}\rho_{j-1}(\r_j)
\end{equation}
the $k$-th order correlation function is evaluated at $\r_1,\dots,\r_k$ as the product of the reduced densities $\rho_{j-1}(\r_j)$. 
To evaluate  the correlation function $\langle \Psi\vert \hat\Psi^\dagger(\r_1)\dots \hat \Psi^\dagger(\r_k)\hat\Psi(\r_k)\dots\hat\Psi(\r_1)\vert\Psi\rangle$ 
the only modification to the single shot algorithm above consists in choosing the 
positions $\r_1,\dots,\r_k$ rather than drawing them randomly.

\subsection*{MCTDHB.}
In the MCTDHB \cite{MCTDHB,Mapping,Package} method the many-boson wave function is expanded in all 
configurations that can be constructed by distributing $N$ bosons
over $M$ time-dependent orbitals $\phi_i(\r,t)$. The ansatz for the time-dependent many-boson wave function reads:
\begin{equation}\label{PsiMCTDHB}
|\Psi(t)\rangle = \sum_{\vec n} C_{\vec n}(t) |\vec n;t\rangle
\end{equation}
In (\ref{PsiMCTDHB}) the $C_{\vec n}(t)$ are time-dependent expansion coefficients
and the $|\vec n;t\rangle$ are time-dependent permanents 
built from the orbitals $\phi_i(\r,t)$.
The MCTDHB equations of motion are derived by requiring stationarity of the many-body Schr\"odinger action functional 
\begin{eqnarray}
 S[\{C_{\vec n}(t)\},&\{\phi_j(x,t)\}] = \int dt \{\langle\Psi(t)|H-i\frac{\partial}{\partial t}|\Psi(t)\rangle \nonumber \\
&- \sum_{k,j=1}^{M} \mu_{kj}(t)[\langle\phi_k|\phi_j\rangle - \delta_{kj}]\}, 
\end{eqnarray}
with respect to variations of the coefficients and the orbitals.
The $\mu_{kj}(t)$ are time-dependent Lagrange multipliers 
that ensure the orthonormality of the orbitals. For bosons interacting via a delta-function interaction and $M=1$ the MCTDHB equations of motion reduce to 
the time-dependent Gross-Pitaevskii equation. 
For more information see the literature \cite{MCTDHB,Mapping,Package}. 

\subsection*{Parameters.}
For the $D=2$ dimensional simulations in this work we assume tight 
harmonic confinement with a frequency $\omega_z$ and a harmonic oscillator length $l_z=\sqrt{\hbar/(m\omega_z)}$ 
along the $z$ -direction. The bosons interact via a regularized contact interaction potential $\frac{\hbar^2\lambda_0}{m}\delta_\epsilon(\r)$, 
with $\delta_\epsilon(\r)=(2\pi\epsilon^2)^{-1}e^{-\r^2/2\epsilon^2}$ 
and a dimensionless interaction strength $\lambda_0=\sqrt{8\pi}a/l_z$, where $a$ is the scattering length and $m$ the mass of  boson. 
We note that it is important to regularize contact interaction potentials for $D>1$ \cite{Greene,Rosti}.
The contributions to the external potential are given by  $V_{x}(x)=\frac{1}{2}m \omega_x^2 x^2$, $V_{y}(y)=\frac{1}{2} m \omega_y^2 y^2$, 
and $V_{g}(x)=C e^{-x^2/2\sigma^2}$, with $C=m\sigma^2\omega_x^2$.
We obtain dimensionless units $\hbar=m=1$ and the Hamiltonian (\ref{Ham})
by measuring energy in units of $\hbar \omega_y$, length in units of $l_y=\sqrt{\hbar/(m\omega_y)}$ and time in units of $1/\omega_y$.
We use a plane wave discrete variable representation to represent all orbitals and operators. 
The width of the contact interaction is $\epsilon=0.15$  and the grid 
spacing is $\Delta x=\Delta y= \epsilon/2$ for all simulations in this work.
For the elongated trap the parameter values are $\omega_x=0.07,\omega_y=1$ 
and $\sigma=10$ on a grid $[-43.2,43.2]\times[-3.6,3.6]$.
For the rotating BEC the parameter values are $\omega_x=\omega_y=1$ and $\eta(t)$ 
is linearly ramped up from zero to $\eta_{max}=0.1$ over a time span 
$t_r=80$. $\eta(t)$ is  then kept constant for $t_{up}=220$ and  ramped back down to zero over a time span 
$t_r$. The potential $V_s(\r,t)=\frac{1}{2}\eta(t)[x(t)^2-y(t)^2]$ rotates harmonically 
with $x(t)= x\cos(\Omega t) + y\sin(\Omega t)$ and $y(t)= -x\sin(\Omega t) + y\cos(\Omega t)$, 
where $\Omega=\pi/4$. The grid size is $[-8,8]\times[-8,8]$.

For the $D=1$ dimensional simulations we assume tight harmonic 
confinement along the $y$- and $z$-directions with a radial frequency
$\omega_\perp=\omega_y=\omega_z$  and an oscillator length $l_\perp=\sqrt{\hbar/(m\omega_\perp)}$. 
The contact interaction potential is then given by $\frac{2 \hbar^2 a}{m l_\perp^2}\delta_\epsilon(x)$, with $\delta_\epsilon(x)=(2\pi\epsilon^2)^{-1/2}e^{-x^2/2\epsilon^2}$. 
We use $\hbar \omega_\perp$ as the unit of energy and $l_\perp$ as the unit length. The dimensionless interaction strength is then given by 
$\lambda_0=2a/l_\perp$. The harmonic potential along the $x$-direction $\omega_x=1/100$ is much weaker than the radial confinement $\omega_\perp=1$. 
The grid size is $[-90,90]$. The Gross-Pitaevskii soliton solution on an infinite line takes on the form $\phi_{mf}(x)=\sqrt{\lambda/4}\sech{(\lambda x/2)}$.\\

\subsection{Image processing.}
The histograms of the positions of particles obtained using the single shot algorithm have a resolution that is determined by the grid spacing.
For better visibility and in analogy to a realistic imaging system we convoluted the 
data points of each histogram with a point-spread function (PSF). As a PSF we used a Gaussian of width 
$3\times3$ pixels.





 
\section*{Acknowledgements}
Financial support through the Karel Urbanek Postdoctoral Research Fellowship is gratefully acknowledged by K. S. 
Computing time was provided by the High Performance Computing Center (HLRS) in Stuttgart, Germany. 

\section*{Contributions}
K. S. and M. K. conceived the ideas and designed the study. K. S. developed the
algorithm and carried out the simulations. K. S. and M. K. wrote the paper.

\section*{Competing Interests}
The authors declare that they have no competing financial interests.

\section*{Corresponding author}
Correspondence and requests for materials
should be addressed to Kaspar Sakmann~(email: kaspar.sakmann@gmail.com)




\end{document}